# Dual Band Receiver Design for the Black Hole Explorer (BHEX) Mission

Cheukyu Edward Tong[1*], Keara Carter[1], Paul Grimes[1], Eugene Lauria[2], Dan Marrone[2], Gabriella Montano[1], Matthew Morgan[3], Yoshinori Uzawa[4] and Lingzhen Zeng[1],

*Abstract*—A dual band receiver has been designed for the Black Hole Explorer (BHEX) mission, which is a space Very-Long-Baseline Interferometry (VLBI) mission concept, aimed at unveiling the photon ring of black holes. The cryogenic receiver comprises a 228-320 GHz Superconductor-Insulator-Superconductor (SIS) receiver, paired with a 76-106.7 GHz HEMT receiver. The details of the design are described in this talk. A novel comb generator, which will be used for delay tracking, has been designed and tested.

## I. Introduction

THE Black Hole Explorer (BHEX) mission [1] aims to reveal the sharp photon ring from light that has orbited black holes --- a phenomenon predicted by the Theory of General Relativity. BHEX will conduct Very-Long-Baseline Interferometry (VLBI) observations in collaboration with ground-based radio telescopes to achieve the unprecedented angular resolution necessary for observing the photon ring. The heart of the BHEX instrument [2] is a dual band cryogenic receiver. The preliminary layout of the receiver was reported at ISSTT-2024 [3]. In this talk, we will describe the design of the BHEX receiver system in further details.

## II. BHEX Instrument Overview

A block diagram of the BHEX science instrument is given in Fig. 1. Several principal subsystems can be identified: (1) a 3.4 m diameter antenna [4], (2) a dual-band receiver system [5], (3) a space cryocooler [6], (4) the digital back end [7], (5) frequency reference [8], and (6) an optical data transmission module [9]. The total analog bandwidth produced by the receiver is 32 GHz, which will be digitized at 1-bit resolution for a total data rate of 64 Gbps. The data will be transmitted to the ground via the optical data transmission subsystem. An ultra-stable quartz oscillator provides the master frequency reference and ensures coherence for tens of seconds.

The two receiver bands are labelled as Rx-L (76-106.6 GHz) and Rx-H (228-320 GHz). Rx-H is the main science receiver and Rx-L, which will be operated at exactly 1/3 of the frequency of Rx-H, is the key to the implementation of the Frequency-Phase-Transfer (FPT) technique [10,11] which is used to extend the coherence of the high frequency receiver, by leveraging the higher signal-to-noise ratio of Rx-L.

BHEX pairs up with ground-based radio telescopes to form the long space-ground baselines needed for high resolution observation of the photon ring of black holes.

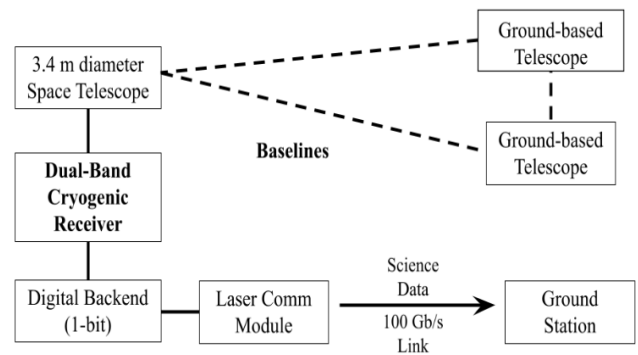

**Fig. 1** Schematic of the BHEX science instrument, showing the principal subsystems as well as the ground-based telescopes with which it form baselines for the space-ground VLBI observation.

## III. Receiver Layout

The beam emerging from the BHEX antenna is relayed to the cryogenic receiver through a beam waveguide [12]. The beams generated by the feedhorns of both Rx-L and Rx-H are designed to have identical far-field patterns. This design enables the use of shaped primary and secondary reflectors to maximize the antenna gain [13]. A set of focusing elements forms the compact beam waists required for the Calibration Module (CALM) as well as the dichroic: CALM provides on-sky beam switching and it can also direct the receiver towards a pair of calibration loads which are thermally anchored to 110 K and 255 K. The dichroic plate splits the beam into two: the low frequency beam is transmitted through it while the high frequency beam is reflected off it. Each of these beams is directed by a focusing mirror to its respective cryogenic receivers through separate openings in the cryostat's radiation shield. Table 1 gives the key design parameters of the dual band receiver system. The two receiver bands will be described in

[1]Center for Astrophysics | Harvard & Smithsonian, Cambridge, MA02138, USA; [2]University of Arizona, Tucson, AZ85721, USA; [3]National Radio Astronomical Observatory, Charlottesville, VA22903, USA; [4]National Astronomical Observatory of Japan, Mitaka, Tokyo, 181-0015, Japan.
*Cheukyu Edward Tong (email: etong@cfa.harvard.edu).

NOTES:

more details in the next 2 sections. Both receivers are configured to operate with dual circular polarizations.

TABLE I. BHEX Receiver at a Glance

|  | Rx-H | Rx-L |
|---|---|---|
| Type | SIS | HEMT |
| Sky Frequency | 228 – 320 GHz | 76 – 106.6 GHz |
| Physical Temp | 4.5 K | 20 K |
| IF | 4 – 12 GHz | 4 – 12 GHz |
| LO Frequency | 240 – 307 GHz | 36 – 55 GHz* |
| Output | DSB | USB or LSB |
| Target Noise Temp. | 30 K | 40 – 50 K |

*Second harmonics mixer used for Downconverter in Rx-L

## IV. SIS MIXER

The heart of the scientific instrument of BHEX is a 225-320 GHz receiver based on the Superconductor-Insulator-Superconductor (SIS) mixer, anchored to the 4.5 K stage of the space cryocooler. A waveguide circular polarizer converts the circularly polarized wave collected by a corrugated feed horn into 2 linearly polarized waves which are separated by a WR-3.4 waveguide orthomode transducer. The receiver will operate in a Double-Side-Band (DSB) mode, to minimize the IF output bandwidth of the receiver, which is set to be 4 – 12 GHz.

The SIS mixer design features a 3-junction series array of coupled to a half-height WR-3.2 waveguide through a single stage thin film microstrip transformer section. The junctions are nominally 1.4 µm in diameter, with a target critical current density of 15 kA/cm$^2$. Devices with barriers of $AlO_x$ and AlN will be explored. The mixer chips are being fabricated in the clean room of NAOJ, Mitaka, Tokyo. Testing of the devices will take place in the coming months.

## V. W-BAND HEMT RECEIVER

The 228-320 GHz SIS receiver is paired with a 76-106.6 GHz W-band receiver based on HEMT amplifier, anchored to the 20 K stage of the space cryocooler. This lower frequency receiver serves as a fringe finder for the main science instrument, and it is also the key to the implementation of the Frequency Phase Transfer (FPT) technique.

The HEMT receiver is also equipped with a corrugated feed horn followed by a circular polarizer coupled to a waveguide orthomode transducer. Two stages of amplification are used for each polarization. The amplifier outputs are sent outside the cryostat with thin wall stainless steel waveguides. A down-converter employs a second harmonic mixer to convert the W-band signals into a 4 – 12 GHz IF. A 76 – 106.6 GHz bandpass filter ensures the down conversion to be Single-Side-Band (SSB).

## VI. COMB GENERATOR

A novel comb generator is included in the receiver system The comb signal will be the basis of operation of the FPT technique, allowing the delay in the receiver signal path to be tracked for both receiver bands. This generator is based on a microwave source, at a frequency $F_0$, close to 10 GHz, which is phase modulated by a 400 MHz sinusoidal signal. The initial narrow comb is frequency multiplied to produce a wider comb to feed a harmonic generator which produces a series of frequency comb signal, centered at $nF_0$, where $n \geq 8$. Since the lower order combs are expected to be much stronger, a waveguide equalizer is used to promote the higher order combs needed for Rx-H. The resulting comb signals will be radiated by a WR-10 horn through a small opening in the hot load assembly of CALM.

A prototype comb generator has been constructed and its output has been observed with both the ngEHT 86/115 GHz receiver [14] as well as the wSMA receiver [15]. A spectrum of the comb is given in Fig. 2.

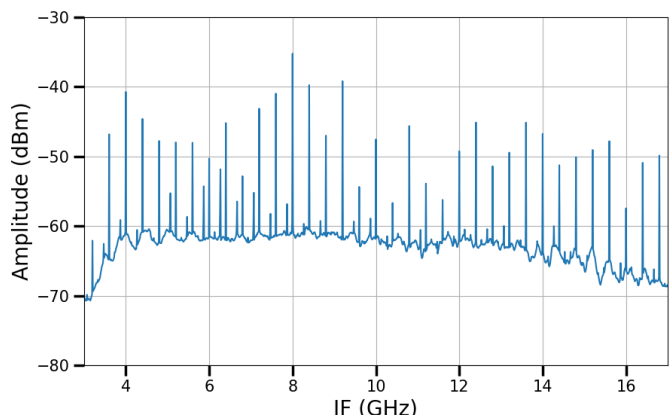

**Fig. 2.** Comb spectrum observed with the wSMA receiver. The base comb is generated with $F_0$ of 10 GHz. The LO of the wSMA receiver is 240 GHz. The main comb teeth are spaced 400 MHz apart.

## VII. CONCLUSION

The receiver design for the Black Hole Explorer mission has entered an advance state. Several prototypes are being developed to ensure that all components demonstrate a high Technology Ready Level (TRL). A novel comb generator has been introduced and tested. The injected comb will be used in the implementation of FTP techniques.

NOTES:

## REFERENCES

[1] M. D. Johnson *et al*, "The Black Hole Explorer: motivation and vision", in *Proc. SPIE, Space Telescopes and Instrumentation 2024: Optical, Infrared, and Millimeter Wave*, 2024, vol. 13092, Art. no. 130922D, Aug. 2024. doi:10.1117/12.3019835.

[2] D.P. Marrone *et al*, "The Black Hole Explorer: Instrument System Overview," in *Proc. SPIE, Space Telescopes and Instrumentation 2024: Optical, Infrared, and Millimeter Wave*, 2024, vol. 13092, Art. no. 130922G, Aug. 2024. doi: 10.1117/12.3019589.

[3] C.E. Tong, P. Grimes, J. Houston, M.D. Johnson, D. Marrone, and H. Rana, "Cryogenic Receiver System for the Black Hole Explorer," presented in the *33rd IEEE Int. Symp. Space THz Tech. (ISSTT2024),* Charlottesville, USA, April 2024.

[4] T.K. Sridharan, *et al*, "The Black Hole Explorer: Preliminary Antenna Design," in *Proc. SPIE, Space Telescopes and Instrumentation 2024: Optical, Infrared, and Millimeter Wave*, 2024, vol. 13092, Art. no. 130926S, Aug. 2024. doi: 10.1117/12.3020504.

[5] C.E. Tong, *et al*, "Receivers for the Black Hole Explorer (BHEX) mission," in *Proc. SPIE, Space Telescopes and Instrumentation 2024: Optical, Infrared, and Millimeter Wave*, 2024, vol. 13092, Art. no. 130926T, Aug. 2024. doi: 10.1117/12.3018054.

[6] H. Rana, *et al*, "The Black Hole Explorer cryocooling instrument," in *Proc. SPIE, Space Telescopes and Instrumentation 2024: Optical, Infrared, and Millimeter Wave*, 2024, vol. 13092, Art. no. 130926U, Aug. 2024. doi: 10.1117/12.3019449.

[7] R. Srinivasan, *et al,* "The Black Hole Explorer: back end electronics," in *Proc. SPIE, Space Telescopes and Instrumentation 2024: Optical, Infrared, and Millimeter Wave*, 2024, vol. 13092, Art. no. 130926V, Aug. 2024. doi: 10.1117/12.3020729.

[8] H. Tomio, *et al,* "Ultra-low-noise laser and optical frequency comb-based timing system for the Black Hole Explorer (BHEX) mission," in *Proc. SPIE, Advances in Optical & Mechanical Technologies for Telescopes & Instrumentation 2024*, vol. 13100, Art. no. 131000O, Aug. 2024. doi: 10.1117/12.3020732

[9] J. Wang, *et al,* "High Data Rate Laser Communications for the Black Hole Explorer," in *Proc. SPIE, Space Telescopes and Instrumentation 2024: Optical, Infrared, and Millimeter Wave*, 2024, vol. 13092, Art. no. 130926W, Aug. 2024. doi: 10.1117/12.3019274.

[10] M.J. Rioja, R. Dodson, T. Jung, and B.W. Sohn, "The power of simultaneous multifrequency observations for mm-VLBI: Astrometry up to 130 GHz with the KVN," *The Astronomical Journal,* vol. 150, p.202, Dec. 2015.

[11] M.J. Rioja, T. Dodson, and Y. Asaki, "The transformational power of frequency phase transfer methods for ngEHT," *Galaxies,* vol. 11, p. 16, Jan 2023.

[12] Ta-Shing Chu, "An imaging beam waveguide feed," in IEEE Transactions on Antennas and Propagation, vol. 31, no. 4, pp. 614-619, July 1983, doi: 10.1109/TAP.1983.1143090

[13] Sridharan, T. K., "The Black Hole Explorer (BHEX): preliminary antenna design", in Space Telescopes and Instrumentation 2024: Optical, Infrared, and Millimeter Wave, 2024, vol. 13092, Art. no. 130926S. doi:10.1117/12.3020504.

[14] C.E. Tong, K. Carter, and L. Zeng, "An 86/115 GHz sidecar receiver – Addition to the ngEHT receiver set for the Owens Valley Radio Observatory and the Large Millimeter Telescope," presented at the 2024 Int. Symp. on Space THz Technology, Charlottesville, VA, April 2024.

[15] P.K. Grimes, R. Blundell, S. Paine, J. Test, C.Y.E. Tong, R.W. Wilson, and L. Zeng, "The wSMA receivers – a new wideand receiver system for the Submillimeter Array," in *Proc. 28$^{th}$ Int. Symp. Space THz Trch.,* Cologne, Germany, March 2017.

NOTES: